\documentclass[conference]{IEEEtran}

\usepackage{url}
\usepackage{cite}
\usepackage{amsmath,amssymb,amsfonts}
\usepackage{algorithmic}
\usepackage{graphicx}
\usepackage{textcomp}
\usepackage{xcolor}
\def\BibTeX{{\rm B\kern-.05em{\sc i\kern-.025em b}\kern-.08em
    T\kern-.1667em\lower.7ex\hbox{E}\kern-.125emX}}

\def\omitauthors #1{{\textit{Authors' name omitted for double-blind review}}}
\def\omitacks #1{{\textit{Acknowledgement omitted for double-blind review}}}

\renewcommand{\baselinestretch}{0.8899}

\begin{document}

\title{Scale up your In-Memory Accelerator: Leveraging Wireless-on-Chip Communication for AIMC-based CNN Inference}

\author{\IEEEauthorblockN{Nazareno Bruschi\IEEEauthorrefmark{1}, Giuseppe Tagliavini\IEEEauthorrefmark{1}, Francesco Conti\IEEEauthorrefmark{1}, Sergi Abadal\IEEEauthorrefmark{2}, Alberto Cabellos-Aparicio\IEEEauthorrefmark{2},}\IEEEauthorblockN{Eduard Alarc\'{o}n\IEEEauthorrefmark{2}, Geethan Karunaratne\IEEEauthorrefmark{3}, Irem Boybat\IEEEauthorrefmark{3}, Luca Benini\IEEEauthorrefmark{1}\IEEEauthorrefmark{4}, Davide Rossi\IEEEauthorrefmark{1}}
\IEEEauthorblockA{
\IEEEauthorrefmark{1}\textit{University of Bologna}, Bologna, Italy, \IEEEauthorrefmark{2}\textit{Universitat Politècnica de Catalunya}, Barcelona, Spain,}
\IEEEauthorblockA{
\IEEEauthorrefmark{3}\textit{IBM Research Europe},
\IEEEauthorrefmark{4}\textit{ETH}, Zurich, Switzerland}
}

\maketitle

\begin{abstract}
Analog In-Memory Computing (AIMC) is emerging as a disruptive paradigm for heterogeneous computing, potentially delivering orders of magnitude better peak performance and efficiency over traditional digital signal processing architectures on Matrix-Vector multiplication.
However, to sustain this throughput in real-world applications, AIMC tiles must be supplied with data at very high bandwidth and low latency; this poses an unprecedented pressure on the on-chip communication infrastructure, which becomes the system’s performance and efficiency bottleneck. In this context, the performance and plasticity of emerging on-chip wireless communication paradigms provide the required breakthrough to up-scale on-chip communication in large AIMC devices. This work presents a many-tile AIMC architecture with inter-tile wireless communication that integrates multiple heterogeneous computing clusters, embedding a mix of parallel RISC-V cores and AIMC tiles. We perform an extensive design space exploration of the proposed architecture and discuss the benefits of exploiting emerging on-chip communication technologies such as wireless transceivers in the millimeter-wave and terahertz bands.
\end{abstract}


\section{Introduction}\label{sec:introduction}
Nowadays, Artificial Intelligence (AI) applications demand more and more computational resources.
The recent advances in the Analog In-Memory Computing (AIMC) research area promise orders of magnitude better peak performance and efficiency over traditional digital signal processing architectures on Matrix-Vector multiplications\cite{sebastian2020memory}, the predominant operations of Convolutional Neural Networks (CNN).

AIMC poses unprecedented pressure on the on-chip communication infrastructure, which becomes the system’s performance and efficiency bottleneck\cite{valavi2019a64tile}.
A fundamental limit of AIMC is technological: the crossbar can not be arbitrarily large (e.g. typically less than 1024$\times$1024 elements); otherwise, the bitlines and the wordlines would become longer and excessively slow and noisy\cite{yan2021uncertaintlymodeling}.
The current limitation on achievable physical dimensions demands a complex mappings strategy to store all the parameters of real-life CNNs, limiting the supported workloads to simple network topologies\cite{wang2020dnnmapping}. Many works like \cite{ottavi2021end2end} have analyzed computational kernels performance of a limited sequence of layers without focusing on system issues.
Researchers have investigated different solutions to meet the requirements of current complex networks\cite{zeng2021mlflashcim, meng2021structuredpruning, zhao2021nnacceleration}, but without detailing how in such complex systems and architectures, interconnections can sustain the many-AIMC throughput\cite{gauchi2019memorysizing}.
Novel approaches tackle this problem by proposing innovative technologies that can complement the wired interconnects with on-chip wireless communication channels\cite{guirado2021dataflow,choi2018onchipcommunication,abadal2020graphenebased}, exploiting the higher bandwidth, versatility, plasticity, and energy efficiency to scale up the number of AIMC devices that can be integrated with sufficient on-chip bandwidth.

In this context, we propose a configurable heterogeneous cluster-based architecture, including programmable RISC-V cores and AIMC accelerators that share a Level 1 (L1) Tightly-Coupled Data Memory (TCDM) at the cluster level and a Level 2 (L2) memory at the system level, combining wired and wireless interconnect technologies.
We evaluate the performance and bottlenecks of the proposed system using benchmarks based on convolution layers. To map the workload on the clusters, we analyze the two main workload distribution approaches: pipelining and data parallelization\cite{narayanan2019pipedream}.
We analyze the benefits of two key wireless communications features, i.e., broadcasting and flexibility, and we estimate up to 5.8 TMAC/s of peak performance (at 350 MHz) and up to 8.2$\times$ of speed-up when we exploit wireless features using a data parallelization workload distribution approach.

\section{Background}\label{sec:background}
This section describes the main technologies we combine and integrate in this work.
\paragraph{AIMC}
In-Memory Computing (IMC) is an emerging computing paradigm exploiting memory arrays delivering orders of magnitude better performance and efficiency than traditional Von Neuman architectures. A computational memory unit is typically organized as a 2-D array, which is referred to as a \emph{crossbar}, with horizontal (wordlines) and vertical (bitlines) wires. The crossbar supports programmable resistors at the cross points of the crossbar, typically featuring a maximum matrix size of 1024 in each of the two dimensions. Many technologies can be used to implement the programmable resistors; in this work, we assume the AIMC core implemented with an array of Phase Change Memory (PCMs) cells\cite{sebastian2020memory}. At the end of each bitline, an analog-to-digital converter (ADC) samples the bitline current and converts it into a digital value.

\begin{figure}[t!]
  \centering
\includegraphics[scale=0.31]{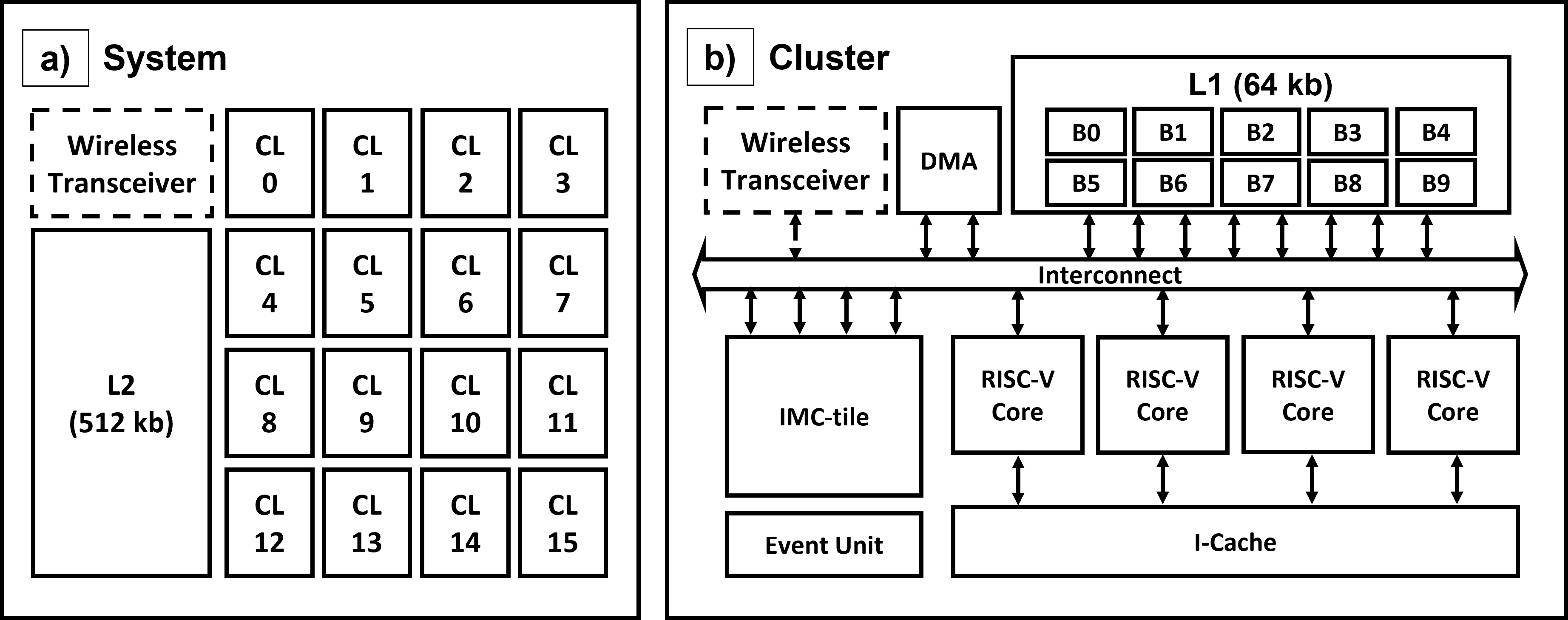}
  \caption{A high-level view of the proposed architecture. (a) Entire system. A set of clusters shares a global L2 memory. (b) Every cluster has its local L1 memory, shared between cores and IMC tile, a DMA, and an event unit.}
\label{fig:architecture}
\vspace{-0.25cm}
\end{figure}

\paragraph{Heterogeneous PULP Cluster}
We use as reference the PULP architecture, an open-source RISC-V based computing platform optimized for embedded parallel and low-power applications.
The PULP Cluster (CL), depicted in Fig.~\ref{fig:architecture}(b), embeds a configurable number of identical RISC-V cores, which share a multi-banked SRAM memory (i.e. L1) accessible from a low-latency logarithmic interconnect. An efficient multi-channel DMA autonomously moves the data from/to L1 to/from L2; a dedicated hardware module, called event unit, handles barriers and other synchronization primitives. L2 resides in a different domain called SoC, connected to the CL through an AXI crossbar. A $256\times256$ analog In-Memory Computing Accelerator (IMA) completes the CL as done in\cite{ottavi2021end2end}.
The IMA streams from and to L1, 8-bit input and output data, and the PCM cells store one's complement 4-bit matrix parameters.
The IMA has 16 4-bytes ports ($IMA_{ports}$), directly connected with L1 through the CL logarithmic interconnect. Data coming from L1 feeds the data path buffers of the accelerator during the \textit{stream-in phase}. Once the input data are ready to be computed, the controller starts the \textit{evaluation phase}, where the output results are converted from  digital-to-analog converters (DACs), computed, and then reconverted from ADCs. The \textit{stream-out phase} stores back in L1 the output data from output ports at the end of the IMA computation.

\paragraph{Wireless}
Recent researches envision wireless networking in the context of heterogeneous multi-chip architectures, based on the integration of antennas and transceivers along with the computing elements of the system and the use of the computing package as the wireless propagation medium \cite{abadal2020graphenebased}. By not requiring to lay down wires between source and destination, the wireless approach bypasses wire routing constraints and offers benefits such as system-wide low latency and scalable broadcast capabilities. Moreover, computing packages are compatible with multiple wideband channels in frequency bands beyond 60 GHz, leading to hypothetical bandwidths in the order of tens or hundreds of Gb/s \cite{chen2019channel}. Finally, such an interconnect provides system-level versatility as bandwidth can be shared dynamically among the antennas adapting to the architecture requirements \cite{guirado2021dataflow}.

\section{System Architecture}\label{sec:system_architecture}\label{subsec:proposed_architecture}

The proposed architecture is depicted in Fig.~\ref{fig:architecture}(a). It is based on a configurable number of CLs ($N_{cl}$), each with a wireless transceiver, an IMA, L1 memory, and a set of RISC-V cores with supporting of DMA and synchronization unit, building a heterogeneous multi-tile system.

\begin{figure}[t!]
  \centering
\includegraphics[scale=0.31]{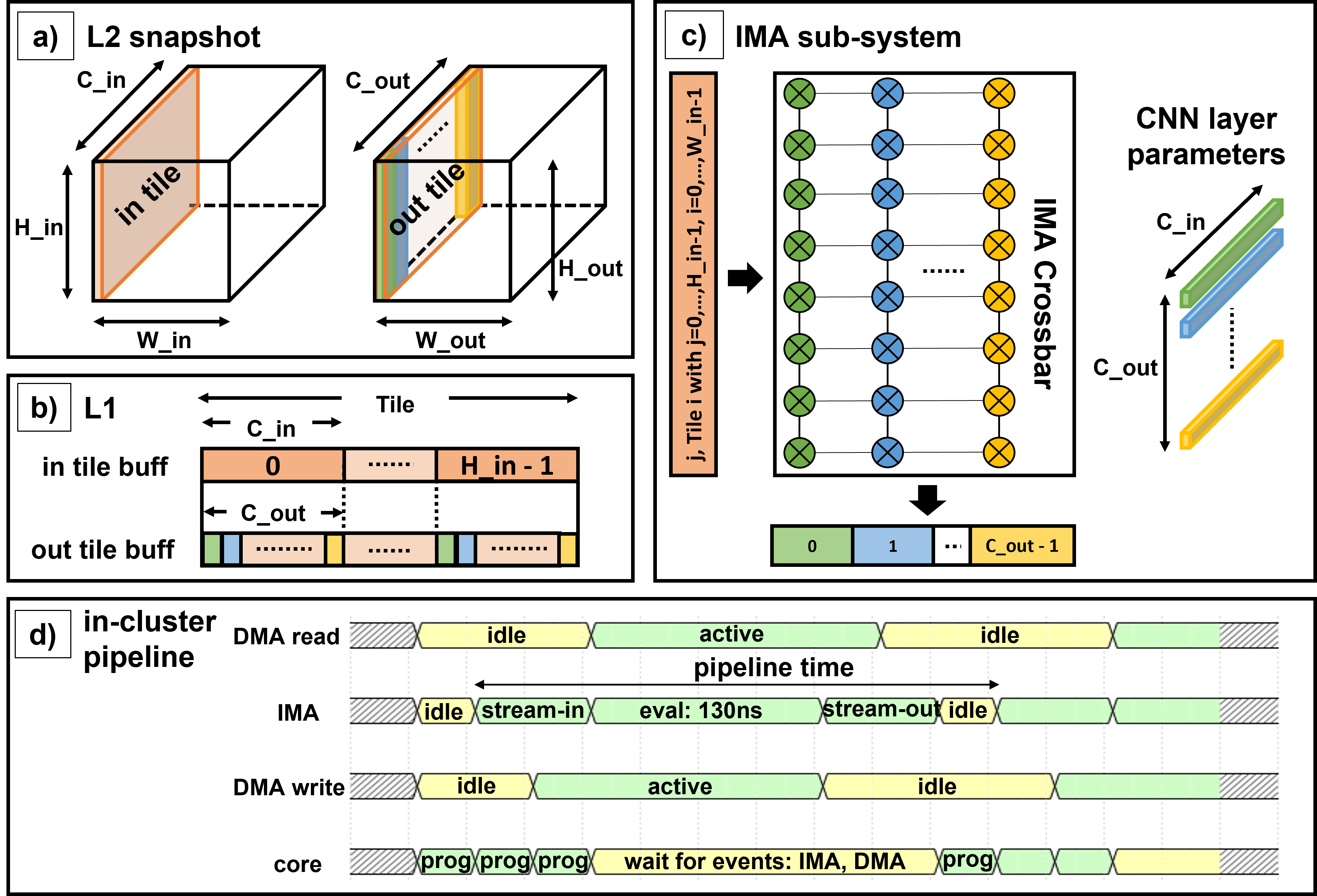}
  \caption{(a) Tiling from and to L2 along W\_in and W\_out dimensions for input and output data respectively (b) L1 input/output tiles buffers arrangement. (c) IMA \textit{stream-in}, \textit{eval} and \textit{stream-out} of a portion of C\_in and C\_out size of the whole input and output tiles. (d) Cluster resources management and synchronization.}
\label{fig:tiling_strategy}
\vspace{-0.25cm}
\end{figure}

On the CL side, the DMA sub-system allows different read/write transfers simultaneously to L1 from L2. It can be programmed by the RISC-V cores of the CL, allowing several outstanding transactions.
The cores can offload a task to the IMA sub-system that autonomously fetches and stores data from L1 and computes the output results, as depicted in Fig.~\ref{fig:tiling_strategy}(c) and described in \cite{ottavi2021end2end}. We can exploit the DMA to tile input data from L2, getting the best performance for the available hardware.
Tiling strategy is crucial to consider a real-life application that requires more than a limited L1 memory space to store inputs and outputs (i.e. 64 kb).
For this reason, we propose a tiling mechanism based on the characteristics of the layers described in Fig.~\ref{fig:tiling_strategy}(a) and (b) and an efficient software-based scheduling to control all the resources of the CL as described in Fig.~\ref{fig:tiling_strategy}(d). In particular, in  Fig.~\ref{fig:tiling_strategy}(d), we can see how the time to create a new context for the accelerator (i.e. prog block in green after the yellow one in the core's waves diagram) is translated into IMA idleness (i.e. throughput reduction), increasing the \textit{pipeline time} and dropping the overall system performance. Moreover, waiting for the events might be longer due to L1 contentions between DMAs and IMA, partially overlapping DMA \textit{active} phases with IMA \textit{stream-in}/\textit{stream-out} ones.

The event unit module efficiently handles the synchronization between CL's resources and among CLs.
It receives hardware events when the DMA, IMA, and cores complete their task. Moreover, the event unit receives software events, adequately programmed by the cores of one CL waiting for the data feed from another CL. The former events synchronize the in-cluster pipeline (i.e., DMA reads, IMA computation, and DMA writes). In contrast, the latter synchronizes the CLs during the multi-CL execution.
The cumulative overhead for synchronization, IMA programming and L1 conflicts is curtailed thanks to the short latency of events managed by the event unit. Sec.~\ref{sec:results} provides a quantitative assessment. 

\section{Workload distribution approaches}
\begin{figure}[t]
  \centering
\includegraphics[scale=0.31]{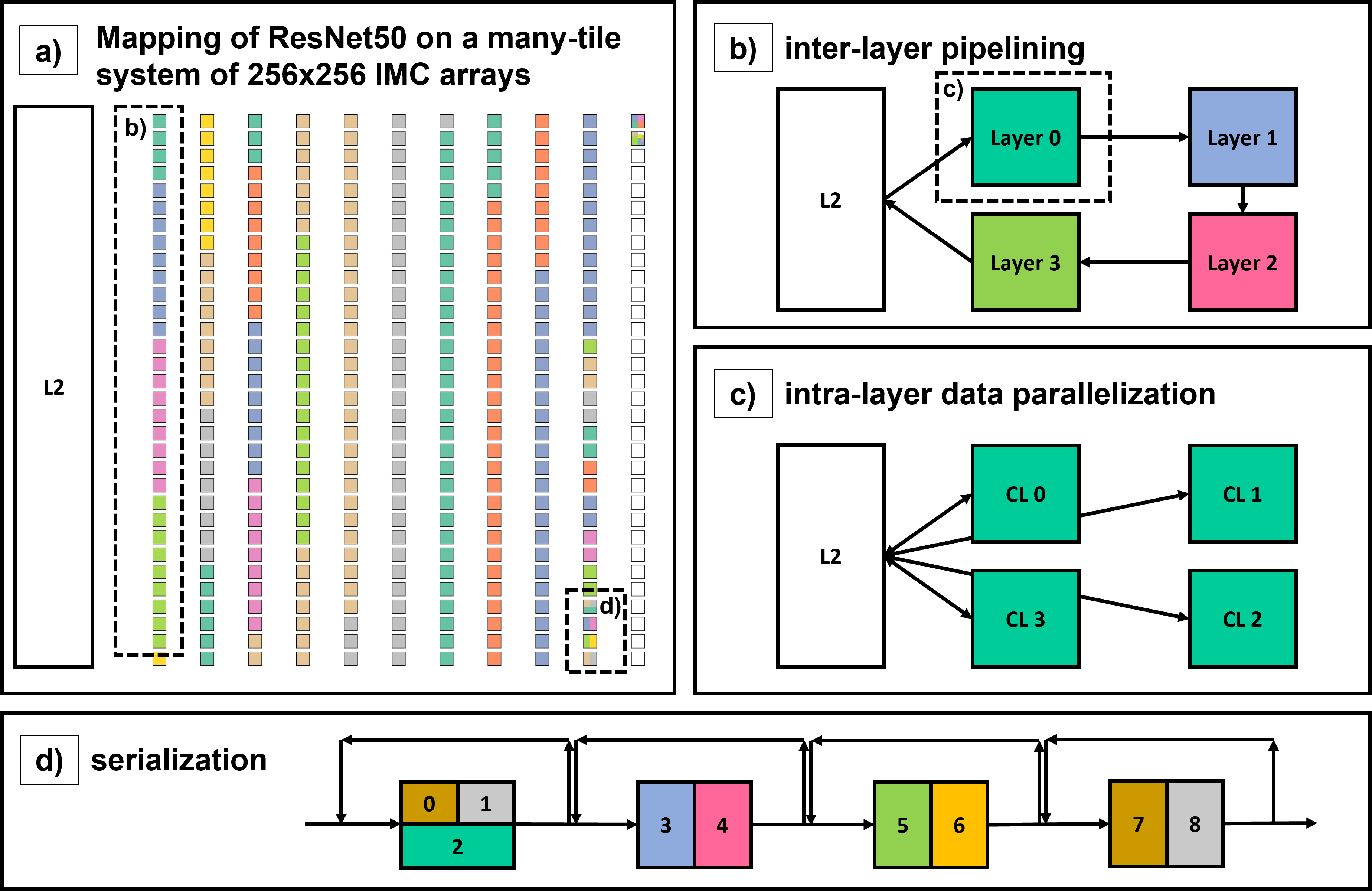}
  \caption{(a) Example of a possible mapping of 33 direct layers of ResNet50. (b) inter-layer pipelining. (c) intra-layer parallelization.
  (d) Layers running on the same AIMC tile have to be executed sequentially, one after each other, as, for example, the last layers which fit just one AIMC tile.}
\label{fig:motivation}
\vspace{-0.25cm}
\end{figure}

We propose two different workload distribution approaches to map the CNN workloads on the multi-CL architecture: \textit{pipelining} and \textit{data parallelization}, motivated by the example of Fig.~\ref{fig:motivation}, where a ResNet50 DNN is mapped on a tiled AIMC-based and requires 322 256$\times$256 AIMC tiles to store the 33 direct layers parameters. In particular, in Fig.~\ref{fig:motivation}(a), every colour represents the mapping of a single layer. Subsequent layers can be performed as a pipeline (i.e. first four layers from the top-right of Fig.~\ref{fig:motivation}(a)). Moreover, many layers must be mapped on different CLs, as, for instance, layer 0 from the top-right of the figure (in teal in Fig.~\ref{fig:motivation}(a), (b) and (c)) is mapped on four different AIMC tiles. Therefore, AIMC tiles dedicated to the same layer can run in parallel.
In the rest of the section, we detail the workload distribution approaches.

\paragraph{pipelining} The \textit{pipelining} workload distribution approach is widely used in AIMC-based architectures\cite{dazzi2021efficientpipelined}, where each layer (or part of a layer) is mapped on an IMA, as shown in Fig. \ref{fig:motivation}(b). The number of pipeline stages is determined by the number of available CLs as well as by the number of layers in the DNN. Every CL computes its output batch using the in-cluster pipeline described in Sec.~\ref{subsec:proposed_architecture}. The first CL fetches the input data from L2, and the last one stores the output results there. The others move data only among L1 memories, as described in Fig.~\ref{fig:motivation}(b). Synchronization between consecutive CLs in the pipeline is performed using a software event that notifies the end of the transaction, awaking the following CL when the compute job is completed. One of the drawbacks of this model is the well-known \textit{pipeline unbalance}. In a pipeline, the throughput is determined by its slower stage. This situation gets worse when multiple layers have to be mapped to the same tile, requiring serialization, as shown in Fig. \ref{fig:motivation}(d), and imposing a trade-off between performance and under-utilization of resources (i.e. IMA).

\paragraph{data parallelization} This workload distribution approach is widely used in many digital architectures such as GPGPUs thanks to its simplicity, well-matched to programming models such as OpenCL and CUDA. On the other hand, non-volatile AIMC-based architectures struggle to exploit this model extensively due to their weight stationary nature. As shown in Fig. \ref{fig:motivation}(c), this model can be exploited when a single layer cannot fit within an IMA. In this case, it can be fully parallelized on multiple IMA. According to this model, a perfect in-cluster pipeline is repeated over $N_{cl}$. Every CL fetches the input data from L2 and computes a portion of output results that are then stored back as described in Fig.~\ref{fig:motivation}(c). One of the drawbacks of this approach is the high pressure on the communication hierarchy, potentially forming a bottleneck at L2, especially when the computational capabilities of the system are significant, as in the case of AIMC tiles.

\section{Simulation methodology}
We model the many-tile system architecture extending the GVSoC \cite{bruschi2021gvsoc} platform, an accurate timing simulator, enabling support for multiple (up to 16) CLs, and extending the interconnect to model conflicts between multiple CLs. As such, in a scenario where \textit{pipelining} among CLs is exploited, only a few conflicts are present because communication is primarily point-to-point among CLs in the  pipeline. Hence, conflicts on on-chip interconnect links can easily be minimized by consecutive mapping layers to directly linked CLs. On the other hand, when a \textit{data parallelization} workload distribution approach is exploited, we expect more conflicts since all the CLs need to communicate with L2.
Since we want to focus on communication effects, we model the L2 memory as a multi-banked scratchpad memory able to sustain the whole bandwidth generated by the system. Only simultaneous addressing of the same memory bank can introduce conflicts within the memory target.

To quantify the limitations of the current state-of-the-art wired interconnects, we propose two interconnection behaviours for every architecture: i) \textit{wired} and ii) \textit{wireless}. In particular, we examine three aggregated interconnect bandwidths between CLs and L2 as \textit{wired}: 22.4, 44.8, and 89.6 Gbit/s at $f_{clock}=350$ MHz, which corresponds to an interconnect bandwidth of 64, 128, and 256 bit/cycle, respectively. In this way, we span a wide range of available wired interconnect resources that can be instantiated in this kind of system\cite{kurth2021anopensourceplatform}. Moreover, we assume a very optimistic latency of 9 cycles between CL and L2.
In the \textit{wireless}, we configured the interconnect bandwidth to 89.6 Gbit/s and reduced the latency to 1 cycle. In this way, we emulate the behaviour of current wireless technology \cite{lee201980}, which is likely to be further improved by novel approaches\cite{abadal2020graphenebased}.
A distinctive characteristic of the wireless interconnect is the seamless support for multicast and broadcast (in our case, from L2 to multiple CLs). In practice, wireless technologies might suffer from packet collisions and losses, which imply re-transmissions, causing a decrease in the effective bandwidth. As detailed above, we model these effects by setting the wireless bandwidth conservatively.

\section{results}\label{sec:results}
We analyze two synthetic benchmarks consisting of only multiply-and-accumulate (MAC) operations, leveraging the maximum throughput of the IMA accelerator. In particular, we consider a sequence of identical 1$\times$1 3D convolutions of 256 input and 256 output channels for the inter-layer \textit{pipelining}, while, regarding the intra-layer \textit{data parallelization}, a single 1$\times$1 3D convolution of 256 input channels and 256*$N_{cl}$ output channels. In both cases, these fit the 256$\times$256 IMA crossbar entirely on each CL, exploiting the maximum throughput available.
We then analyze the results considering several CLs configurations, mappings, and bandwidths.
For all the assessments, we use a computation efficiency metric $\eta$ based on the theoretical limit on the obtainable GMAC/s by the $N_{cl}$ available IMAs with $T_{eval}=130$ ns of analog eval time each. We call it \textit{baseline}. It corresponds to $$baseline = \frac{10^{-9}*N_{cl}*C_{in}*C_{out}}{T_{eval}+T_{\operatorname{\mathit{stream-in}}}+T_{\operatorname{\mathit{stream-out}}}}\quad \text{[GMAC/s]}$$
where $T_{\operatorname{\mathit{stream-in(out)}}}=\frac{1}{f_{clock}}*\frac{C_{in(out)}}{IMA_{ports}*4}[s]$.
We report how efficient the total execution in every scenario, extracted from GVSoC simulation (i.e. $tot\_exec\_cycles$) is with respect to the \textit{baseline} (i.e. $\eta$), corresponding to $$\eta = \frac{10^{-9}*f_{clock}*\frac{N_{cl}*C_{in}*C_{out}}{tot\_exec\_cycles}*100 }{baseline}\quad \text{[\%]}$$

\begin{figure}[t!]
  \centering
\includegraphics[scale=0.31]{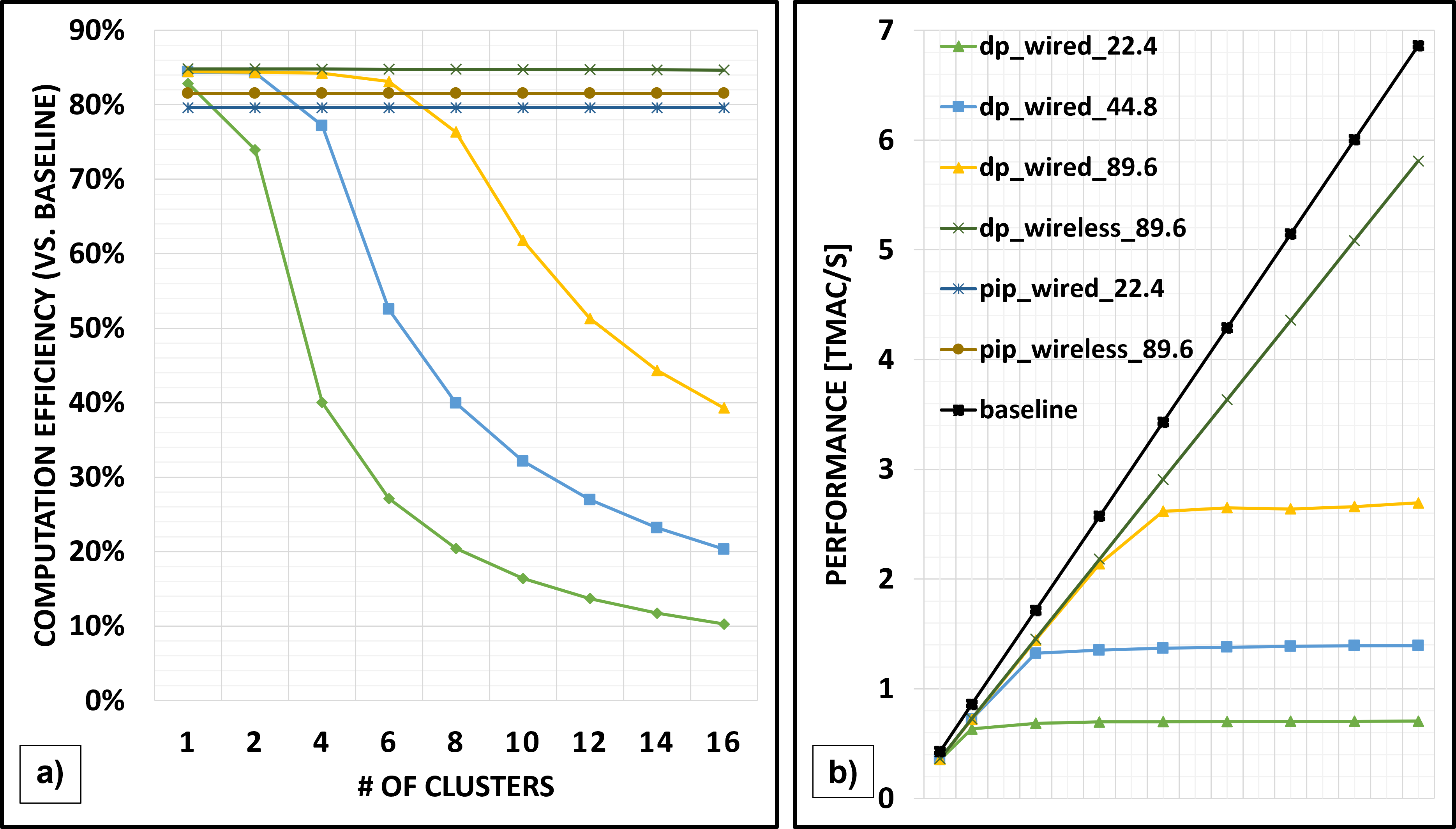}
  \caption{(a) Computation efficiency $\eta$ of both mappings, increasing the number of CLs and comparing \textit{wired} and \textit{wireless} configurations. (b) The peak of performance in data parallelization workload distribution comparing bandwidths and technologies.}
\label{fig:comparison}
\vspace{-0.25cm}
\end{figure}

Fig.~\ref{fig:comparison}(a) shows the comparison between two workload distribution approaches in both scenarios, \textit{wired} and \textit{wireless}, when we increment $N_{cl}$ and the \textit{wired} bandwidth. This can only be achieved in the case of wired communication with complex and area-expansive topologies such as mesh NoCs, which are well known not to scale with the number of CLs. We can see how two workload distribution approaches reach a very high level (e.g. 80\%) of computational efficiency in a single-CL execution. The gap between this result and the ideal case (i.e. 100\% efficiency) is due to the contentions in L1 between IMA and DMA and the software programming overheads.

Moving to the more realistic case of multi-CL execution, starting from the inter-layer \textit{pipelining} model, we see a constant trend in the computation efficiency due to our scalable runtime. Here, the benefits of having higher bandwidth for the computation efficiency are irrelevant since data transfers are overlapped with IMA computation, and communication is never the bottleneck for performance. Still, the lower latency of the wireless links reduces the time one CL needs to wait for input data by 2\%.

On the other hand, the intra-layer \textit{data parallelization} model has completely different behaviour. In the \textit{wired} curves, we can see how computation efficiency struggles with the conflicts of large multi-CL systems contending the same communication resources even if pushing the aggregated interconnect bandwidth to the same of the \textit{wireless}. This happens because every CL simultaneously tries to fetch the same input data of the other CLs from L2 and then writes back the output results. In the \textit{wireless} scenario, instead, the broadcasting features can be exploited, reducing to zero the conflicts on the communication channel. For this reason, the computation efficiency is improved up to 8.2$\times$, 4.1$\times$ and 2.1$\times$ respectively for 22.4, 44.8, and 89.6 Gbit/s.

In Fig.~\ref{fig:comparison}(b), we evaluate the effective performance of our system by analyzing the interconnect bandwidths of \textit{wired} architecture with respect to the \textit{wireless} communication methodology that we propose, and we can see a linear trend up-scaling the AIMC tiles and up to 5.8 TMAC/s using the \textit{wireless}.

\section{Conclusion}\label{sec:conclusion}
We presented an AIMC-based multi-tile heterogeneous architecture, analyzing the performance peaks when computing typical CNN workloads and providing insights about the limitations caused by the limited bandwidth of classical communication channels, their rigidity, and the physical size of the analog devices. In this context, the performance and plasticity of emerging on-chip wireless communication paradigms can provide a solid solution in terms of computation performance, especially considering real-life applications with the need of splitting the computation along different AIMC. Open challenges are still balancing the different layers workloads, parallelizing the slowest layers to reduce the pipeline time (i.e. critical path).

\section*{Acknowledgments}
This work was supported by the WiPLASH project (g.a. 863337), founded from the European Union’s Horizon 2020 research and innovation program.

\bibliography{bibliography}
\bibliographystyle{ieeetr}

\end{document}